# Optical detection of structural properties of tumor tissues generated by xenografting of drug-sensitive and drug-resistant cancer cells using partial wave spectroscopy (PWS)

Prakash Adhikari,[1] Prashanth K. B. Nagesh,[2] Fatimah Alherthi,[1] Subhash C. Chauhan,[2] Meena Jaggi,[2] Murali M. Yallapu,[2*] and Prabhakar Pradhan[1*]

[1]Department of Physics and Astronomy, Mississippi State University, Mississippi State, MS, USA, 39762
[2]Department of Microbiology and Immunology, School of Medicine, University of Texas-Rio Grande Valley, McAllen, Texas-78504

*E-mails: PPradhan: pp838@msstate.edu  and  MMYallapu: murali.yallapu@utrgv.edu

The quantitative measurement of structural alterations at the nanoscale level is important for understanding the physical state of biological samples. Studies have shown that the progression of cancer is associated with the rearrangements of building blocks of cells/tissues such as DNA, RNA, lipids, etc. Partial wave spectroscopy is a recently developed mesoscopic physics-based spectroscopic imaging technique which can detect such nanoscale changes in cells/tissues. At present, chemotherapy drug treatment is the only effective form of treatment; however, the development of drug-resistant cancer cells is a major challenge for this treatment. Earlier PWS analyses of prostate cancer cells, a 2D structure, have shown that drug-resistant cancer cells have a higher degree of structural disorder compared to drug-sensitive cancer cells. At the same time, structural properties of the metastasize tumor grown to 3D structure from drug-resistant and drug-sensitive cancer cells within the body is not well studied. In this paper, the structural properties of tissues from grown 3D tumors, generated from docetaxel drug-sensitive and drug-resistant prostate cancer cells xenografted into a mouse model, are studied. The results show that xenografted tumor tissues from drug-resistant cells have higher disorder strength than the tumor generated from drug-sensitive prostate cancer cells. Potential applications of the technique to assess chemotherapy effectiveness in cancer treatment are discussed.

**Keywords:** *Light Scattering, Prostate Cancer, Xenograft, Disorder Strength, Docetaxel, Spectroscopy*

## 1    Introduction

Elastic scattering, especially in the visible range of light, is an important method for probing structural morphologies of the biological cells/tissues. It is now shown that probing the structural alteration at nano to submicron scales enables the prediction of several properties of the physical conditions of cells/tissues. In addition to this, the progress of carcinogenesis results in nanoscale structural alteration due to the rearrangement of macro molecular components inside the cells/tissues. This nanoscale structural alteration is considered an important biomarker in the determination of cancer stages, as well as to assess the efficacy of a chemotherapy drug at the different levels of tumorigenicity [1–3]. However, the histopathological examinations of cells/tissues, conventionally, are based on a large degree of changes in the cellular architecture during the diseases process [4]. Also, the sensitivity of the existing optical microscopic techniques used to detect such nanoscale alterations are restricted by the diffraction limited resolution (>~200nm).

A recently developed spectroscopic microscopy technique, partial wave spectroscopy (PWS), which combines the interdisciplinary approaches of mesoscopic condensed matter physics and microscopic imaging, is used to quantify the change of the nanoscale structural disorder of weakly disordered biological medium like cells/tissues [3,5]. The statistical quantifications of the reflected intensities due to the nanoscale refractive index fluctuations of the biological cells/tissues are carried out using the PWS analysis. In the PWS technique, the backscattering signals from thin weakly disordered cell/tissue samples are divided into many parallel scattering quasi one-dimensional reflections to calculate the structural disorder strength of the samples[3,5]. Further, the spatial variation of the intracellular components such as DNA, RNA, lipids, and extracellular matrices (ECM) gives rise to spatial mass density fluctuations in terms of the refractive index fluctuations of the cells/tissues [6,7]. This spatial refractive index fluctuations can be quantified in terms of the degree of structural disorder. The degree of structural disorder parameter $L_d$, called the disorder strength, can quantify nanoscale alteration and distinguish different cancer stages accurately. Here disorder strength $L_d = dn^2 \times l_c$, where $dn$ is the standard deviation (*std*) of the onsite refractive index fluctuations $dn(r)$ and $l_c$ is its spatial correlation length.

The efficiency and application of the PWS technique in measuring nanoscale alteration, i.e. the $L_d$ parameter, to diagnose diseases like cancer has been developed and explored [5,7–9]. It is further established using the cancer cell lines that drug-sensitive and drug-resistant cancer cells have different structural disorders. Drug-resistant cancerous cells are able to survive chemotherapy drug treatment due to the different mechanisms responsible for drug resistance and the development of different morphological structures. These different morphological structures in drug-resistant cells may be due to the rearrangements of macromolecules, increase in the sizes of pores, architectural differences of cytoskeletal network, etc. which result in increasing aggressiveness which in turn increases the disorder strength. PWS technique was successfully used to study the effect of chemotherapy drugs on cancerous cells and to quantitatively measure the disorder strength of drug-sensitive and drug-resistant cancerous cells [10].

Prostate cancer is one of the most prevalent types of cancer with the highest male mortality rate in the USA. The American Cancer Society (ACS) estimates about 174,650 new cases of prostate cancer will appear and account for a total of 31,620 deaths for 2019. Across the globe, the statistical data of prostate cancer suggests that among every 9 men, one individual will develop this cancer during his lifespan. Therefore, it is necessary to explore early and effective diagnosis/treatment methods for prostate cancer. At present, chemotherapy is the only way to treat metastasized prostate cancer, however, it is often found ineffective due to an individual patient's chemo-resistance that leads to tumor progression [10–13]. The PWS studies of cancer cell lines have shown some promising success in distinguishing the hierarchy and drug effectiveness based on the disorder strength $L_d$ parameter, however, these studies were mainly focused on backscattering signals from isolated cancer cells where the cells were grown in 2D on glass slides. In reality, a metastasized cancer cell grows into a tumor with 3D structure when it grown within the body, and these tumor cells may have different structural properties due to its 3D growth into tissue structures. This leads to a demand for the development and characterization of 3D tumor tissues that are generated from the drug-sensitive and drug-resistant cancer cells. This could establish a correlation between the isolated cells grown in 2D and the cells grown in a tissue in 3D based on the structural parameter using the PWS technique. Human cancers have been studied by innumerable murine methods and the determinants responsible for malignant transformation, invasion and metastasis, as well as the examination of response to therapy is investigated by the aid of these murine models. At the same time, the structural disorder properties of cancer growth are not well studied by the xenografting of cancer cells. Therefore, study of cancer stages and drug effect on cancer tissues that are grown into a 3D tumor from different types of metastasized cancer cells using PWS technique will provide a connection between 2D *in vitro* cell growth to 3D *in vivo* tissue growth of the same types of cells.

In this work, using the PWS technique, we explore the structural properties of the 3D tumor tissues obtained by xenografting drug-sensitive and drug-resistant human prostate cancer cells in a mouse model. We have studied structural properties of tissue obtained by xenografting two types of human prostate cancer (PC) cell lines, namely DU145 and PC-3, whose drug-resistant and drug-sensitive structural properties were studied earlier by PWS technique using $L_d$ parameter [3,5,10]. The earlier PWS results from prostate cancer cell lines showed an increase in the aggressiveness or tumorigenicity for drug-resistant cancer cells, relative to its drug-sensitive cancer cells. The cells that grow on a slide are mainly 2D in nature, and have shown above trend. In particular, here we want to verify the structural properties of tissues based on disorder strength or $L_d$ value when they grow in 3D structure by xenografting these cells, and to understand any relationship with their original 2D cells. Xenografting of human cancer cells in a mouse model is one of the most extensively used models to study the development of tumors from cells [14]. Cancerous human cells were subcutaneously injected in immunocompromised mice. Based on the number of cells injected, the tumors will develop over 1-8 weeks and reaction to the proper therapeutic regimes can be studied *in vivo* [14,15] or *ex vivo*. The disorder strength $L_d$ of 4µm thickness excised tumor tissue were calculated, and the correlations of structural disorder values of these xenografted tissues corresponding to their original cells are compared. This study may provide the structural disorder difference of 2D cell lines and their corresponding growth of 3D structures and any correlations. Therefore, this new PWS analysis of human tumor cell xenografting can be used to study the physical state and effectiveness of a chemotherapy drug in cancerous cells/tissues to improve chemotherapy treatment methods.

## 2 Method

### 2.1 PWS Experimental Setup:

We perform the structural disorder measurement using a recently developed partial wave spectroscopy (PWS) experimental technique, with added further engineering of finer focusing. The partial wave spectroscopy (PWS) setup with a fine focus to measure the structural alterations at nanoscale level is as shown in Fig.1. Xenon Lamp (Newport, 150W), a source of stable broadband white light is used to illuminate tissue samples of micron thickness using Kohler Illumination. The white light is reflected towards the combination of lenses with silver coated mirror (Thorlabs, f=50.8mm). The combination of converging lenses (Thorlabs, f=50.8mm) along with the apertures (Newport) form a 4f system that helps to minimize the diffraction effect and preserve the high-frequency effect and hence enrich the sharpness in an image. This collimated light from the lens is reflected by a right-angle prism (Thorlabs, 25.4mm) and passed through the dichroic mirror (Thorlabs, 25.4mm) and then enters an objective lens (Newport, NA=0.65,40X). The low numerical aperture objective lens focuses the light in the sample within its working distance with the help of high-resolution 3D electronic motorized stage (Zebar Tech, 100nm in Z axis and 40nm in X-Y). This high-resolution motorized 3D stage is considered revolutionary to the microscopic setup for its extreme accuracy and finer focus. A finer focus is essential for correctly defining the effective scattering volume/length of a sample. The backscattered signal from the sample is passed through the objective which gets reflected into the thick collecting lens (Thorlabs, Φ=50.8mm) before the aperture. Finally, the collected backscattered signal passes through a liquid crystal tunable filter i.e. LCTF (Thorlabs, KURIOS-WB1) with spectral resolution of 1nm within the visible range (420-730nm) of the spectrum. A CCD camera (Retiga 3, 1460×1920) coupled with a LCTF controller records the filtered signal in the desired wavelength range. Here, the CCD camera and LCTF filter are coupled with the LCTF controller in such a way that for each 1nm increment in wavelength by the LCTF filter (resolution 1nm), the backscattered signals are recorded in the CCD within the visible range.

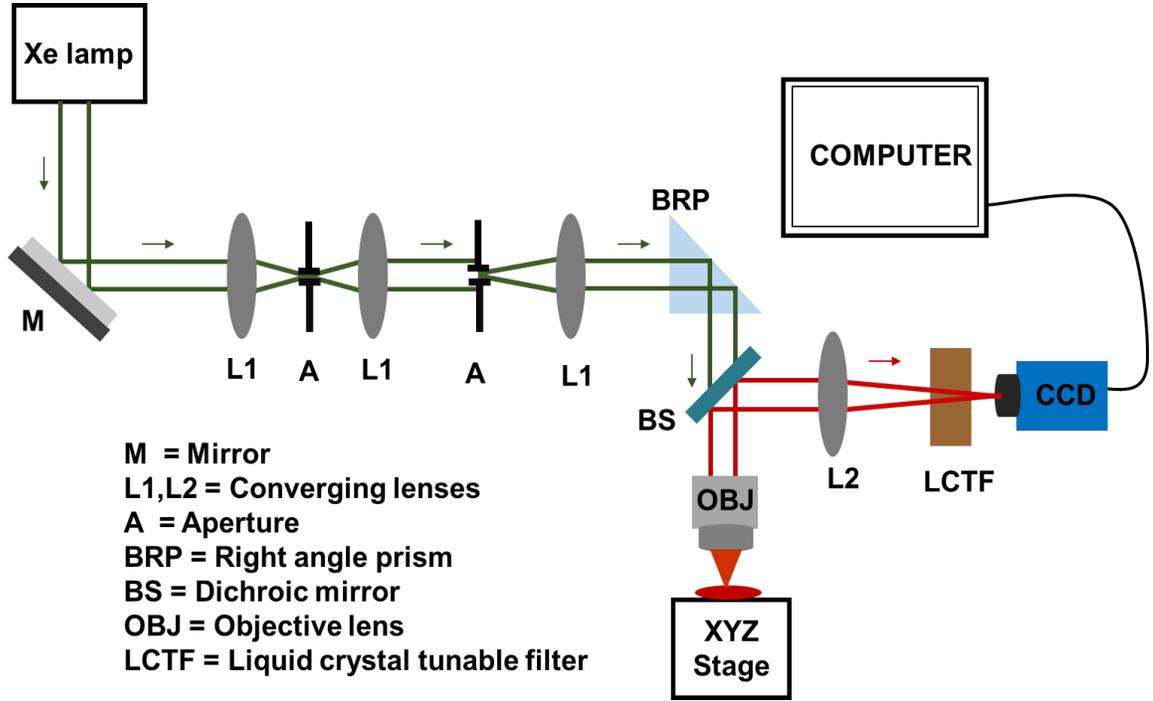

**Fig.1** Experimental layout of the Partial wave spectroscopy (PWS) setup with nanoscale sensitivity. The green color represent broadband white light from the source and red color after dichroic mirror is the back scattered signal.

2.2 *Calculation of the structural disorder or disorder strength $L_d$:*

The backscattered images are recorded in the CCD camera at every wavelength *($\lambda$)* at the spatial pixel position *(x,y)* in the wavelength range 450-700nm, and the reflected data cube, *R(x, y; $\lambda$),* is acquired by PWS system. Here, the data cube *R(x,y;$\lambda$)* includes the fluctuating part of the reflection coefficient over the visible wavelength regime due to the presence of a disordered medium including the high frequency noise. In a quasi-1D approximation, the collected backscattered data at each *(x,y)* from *R(x,y;$\lambda$)* is fitted with a polynomial of the 5$^{th}$ order. The fitted polynomial is then extracted from the signal to remove the systematic errors. In the next step, the *R(k)* signal for each pixel position *(x,y)* is obtained after applying a fifth-order low-pass Butterworth filter with a suitable normalized cutoff frequency to remove the high frequency noise components from the reflected signal of micron size samples. From the extracted backscattered data cube, calibration of the system is done by matching the reflected intensity pattern using a non-disordered system, NIST-traceable microspheres with reflected signal from a thin film slab model. PWS measures the spectral fluctuation dividing the signal from the sample into the collection of backscattered parallel channels. Based on the mesoscopic optical property of the object, the spectral fluctuations originating from multiple scatterings of the sample are analyzed.

The fluctuating part of the reflection signal $R(x,y;\lambda)$ arises due to the multiple interferences of the photons reflected from the disordered medium. Since in a quasi-1D approximation, the sample is virtually divided into many parallel channels within the diffraction limited transverse size and the backscattered signal propagating along the 1D trajectories are collected. This optical detection method is termed as the partial wave spectroscopy (PWS). The statistical properties of nanoarchitecture are quantified at the nanoscale level by analyzing the fluctuating part of the reflected intensity. The refractive index fluctuation information is collected from these spectral fluctuations originated from the multiple scattering of the sample at any length scales below the diffraction limit. The degree of structural disorder parameter $L_d$ can be derived from the *rms* value of the reflection intensity $<R(k)>_{rms}$ and the spectral auto-correlation decay of the reflection intensity $C(\Delta k)$ ratio. That means, for a given pixel at position *(x,y)*, the degree of structural disorder is defined as [3,5]:

$$L_d = \frac{B \langle R \rangle_{rms}}{2k^2} \frac{(\Delta k)^2}{-\ln(C(\Delta k))}\bigg|_{\Delta k \to 0}. \qquad (1)$$

Where $B$ is the normalization constant, $k$ is the wave number ($k = 2\pi/\lambda$).

For the Gaussian color noise of the refractive index at position *r* and *r'*, $<dn(r)dn'(r')>=dn^2 exp(-|r-r'|/l_c)$, it can be shown that $L_d=<dn^2>l_c$ [3,5], however, this form may change due to different situations. The disorder strength quantifies the variability of local density of intracellular material within the samples, and hence the average and standard deviation of the $L_d$ are calculated to characterize the system.

2.3 *Sample preparation:*

Freshly collected prostate cancer (PC) cells from two different cell lines were used to develop prostate tumor using Xenografting in mouse model. For the generation of tumor xenograft mouse models, 6–8 weeks old male nude mice were implanted with PC-3 (docetaxel-sensitive and docetaxel-resistant) and DU145 (docetaxel-sensitive and docetaxel-resistant) human PC cell samples ($2 \times 10^6$ cells per mice), by subcutaneously injections. After tumors reached beyond the critical size of 1000 mm³, they were excised from euthanized mice. The excised tumors were further paraffin embedded and sectioned using microtome of 4 μm thickness and placed on glass slides. Further, these slides were processed for antigen retrieval process as described previously [16]. The resultant tumor sections were subject for imaging studies.

## 3   Result

PWS detects the nanoscale structural alteration in the cells/tissues and can distinguish the different levels and effect of drug in the tumorous cells/tissues [7,10]. Among the different types of cancer, prostate cancer is a major concern of public health at present because of its low survival rate. Further, drug-resistant cells are a prominent problem currently in cancer treatment. Therefore, we focus our research to characterize the structure of tumor tissues generated by the xenografting of chemotherapy drug-sensitive and drug-resistant prostate cell lines, DU145 and PC-3. For this, human drug-sensitive and drug-resistant prostate cancer cells were subcutaneously injected in mice and allowed to grow and achieve the tumor size of ~1000mm$^3$. After that mice were euthanized, and tumors were excised then subsequently processed to 4 μm thick tumor sections on slides for PWS imaging analyses.

PWS experiments, as described earlier, were performed on xenografted prostate tumor tissues for each category on at least 3 different mice. All tumor tissue was obtained from different prostate cancer cell lines xenografted into mouse models. For each tissue, PWS experiments were performed on 7 different spots. Therefore, for each category of tumor, ~60 different spots are experimented with PWS and analyzed. The spectroscopic PWS experiments are performed in the wavelength range 450-700nm. The backscattered data matrix *R(x, y; λ)* are imported and the disorder strength for the each tissue is calculated using the PWS technique as describe in the section 2.2. The disorder strength is calculated as the product of the variance and the spatial correlation length of the refractive index fluctuations, *$L_d=<\Delta n^2>l_c$*. The average and standard deviation of the disorder strength for each category is calculated in order to understand the physical properties of a tumor developed as a 3D structure from a cell. The detailed PWS analyses of tumors obtained from xenografted model of drug-sensitive and drug-resistant cell lines DU145 and PC-3 are explained below:

### 3.1 Structural Disorder in the xenografted DU145 tumor tissue type:

In Fig. 2, the PWS analysis of a tumor obtained from xenografting drug-sensitive and drug-resistant xenografted prostate human cancer cell line of DU145 type are shown. From the PWS experiment *R(x,y,λ)* data matrices were obtained. At every pixel point *(x,y)*, *$R(k)_{rms}$* value and corresponding $C(\Delta k)$ were obtained, and from these two values, $L_d$ value was calculated using Eq.(1). The bright field images of the thin tissue samples developed from drug-sensitive and drug-resistant cancer cell lines appear indistinguishable, whereas the $L_d$ images are noticeably distinguishable. The red spots in the $L_d$ image represent a higher disorder strength, i.e. the $L_d$ value of that pixel. It can be seen in the bar graphs that there is an increase in the degree of the structural disorder of tumor tissue generated from xenografting drug-resistant cancer cells, compared to the tissues obtained from xenografting drug-sensitive cancer cells. The average $L_d$ value of tumors obtained from drug-resistant cancer cells is 9% higher than the tumors obtained from drug-sensitive cancer cells and corresponding standard deviation $std(L_d)$ is 8% higher. This

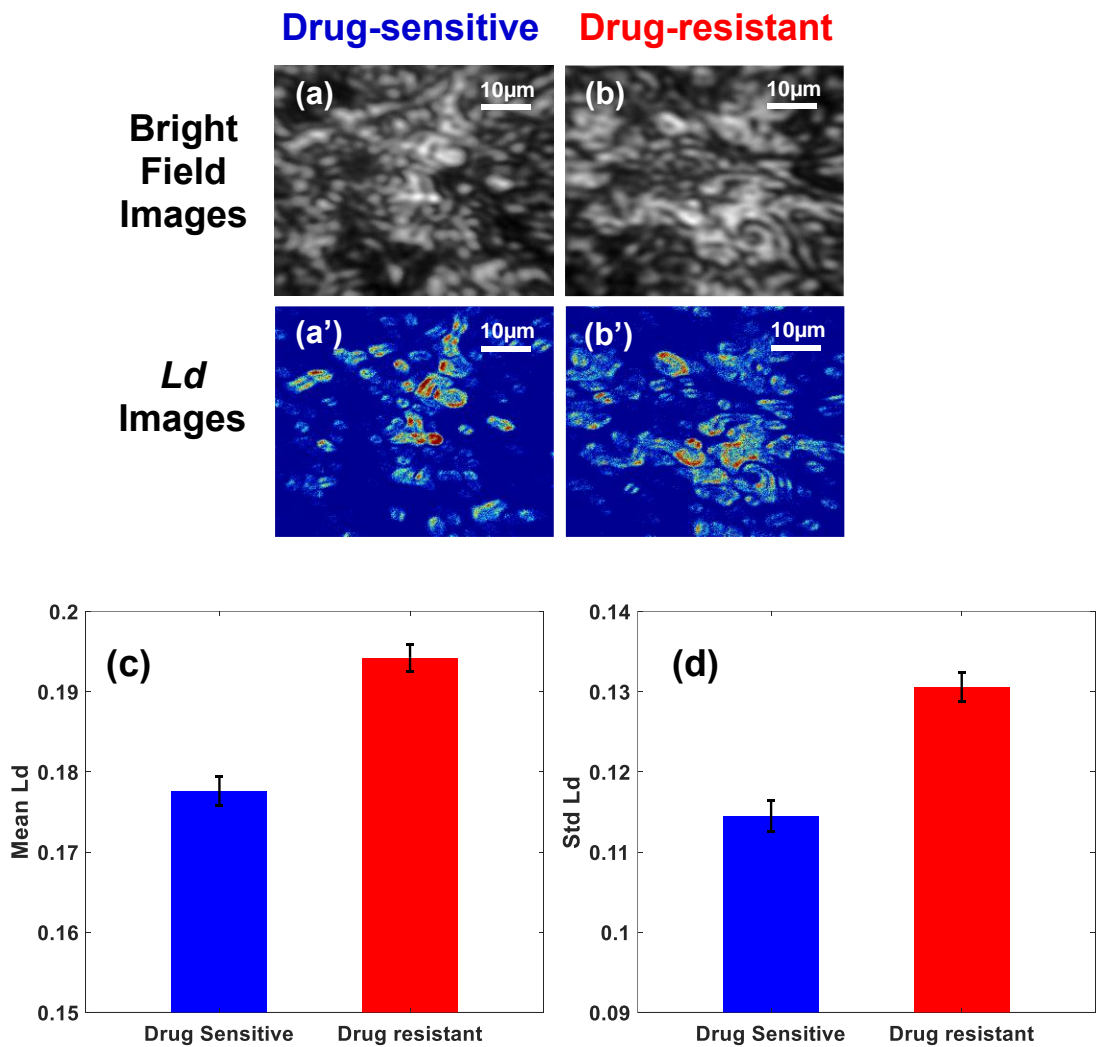

**Fig. 2** (a) and (b) are the bright field images while (a') and (b') are the disorder strength, $L_d$ images of tissue obtained by xenografting drug-sensitive and drug-resistant PC cells line of DU145 type respectively. (c) and (d) are the graphical representation of mean and standard deviation of $L_d$ of tumor respectively. Result shows 9% higher in mean $L_d$ and 8% increase in $std\ L_d$ in tumor developed from drug-resistant PC DU145 type cells line than the drug-sensitive cells. P-value < 0.05.

result is in strong agreement with the disorder strength calculated for drug-resistant and drug-sensitive cell lines earlier [10]. The disorder strength of the prostate cancerous cell line calculated using PWS have shown that, the average and standard deviation of $L_d$ values are higher in drug-resistant cells compared to the drug-sensitive for DU145 type cells. That means the xenografted tissue structure also have similar trends to original cell structures.

*3.2 Structural Disorder in the xenografted prostate PC-3 tumor tissue type:*

Fig 3. shows the bright field and $L_d$ images of tumor tissues originated from drug-sensitive and drug-resistant PC PC-3 type cells. Based on the intrinsic properties of the tissue, the disorder strength $L_d$

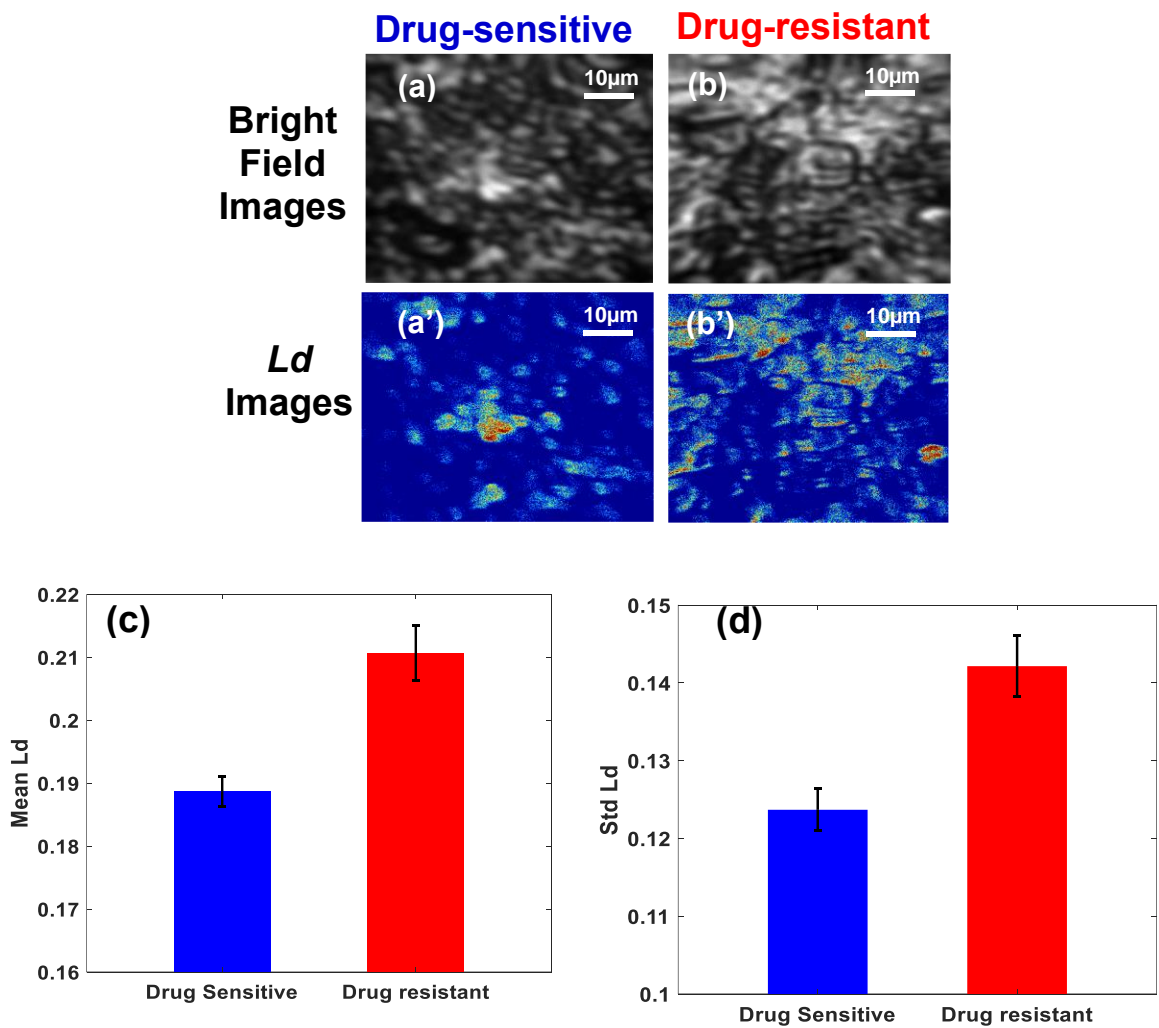

**Fig. 3** (a) and (b) are the bright field images and (a') and (b') are the disorder strength, $L_d$ images of tissues obtained by xenografting of drug-sensitive and drug-resistant PC cells line of PC-3 type respectively. (c) and (d) are the bar graph representation of mean and standard deviation of $L_d$ of tumor respectively. Result shows 12% higher in mean $L_d$ and 15% increase in *std* $L_d$ in tumor developed from drug-resistant PC PC-3 type cells line-than the drug sensitive cells. P-value < 0.05.

at each pixel of the individual tissue image is calculated and represented by 2D color maps. In the color map, red spots correspond to the higher structural disorder strength present in the thin tissue structure (averaged along the z-direction of the sample). The bar graphs are the representation of the average and standard deviation of $L_d$ values with the standard error bars. The result shows 12% difference in the average local disorder strength $L_d$ and 15% difference in *std* $L_d$, between tumors obtained from drug-sensitive and drug-resistant cell lines. The increment in mean and standard deviation of $L_d$ for tissues collected by the xenografting of drug-resistant PC PC-3 cells compared to the drug-sensitive cells are consistent with the original cell structures.

It is clear from the plots that the disorder strength increases form drug-sensitive to drug-resistant tumor tissue samples. Earlier results using PWS analysis show that PC PC-3 cells are more aggressive than other cell lines and in the same way drug-resistant PC cell lines have higher disorder strength than drug-sensitive cells. Figs 2 and Figs 3 show the results obtained using xenografted tissue samples also have the same kind of disorder strength hierarchy to that of the original PC cells. In particular, a comparatively higher structural disorder strength $L_d$ for tumors obtained from drug-resistant PC-3 PC cell line than DU145 PC cell line is in strong agreement with their original PC cell line disorder strength. This confirms that the degree of disorder strength $L_d$ can be used as a marker to detect the cancer stages or drug effects using 3D cancer tissues structure, similar to that of a cell line that is easy to study.

## 4   Conclusions

*The result:* In this work, we have reported the PWS study of the nanoscale structural changes in tissues generated from the xenografting of drug-sensitive and drug-resistant PC cells using a mouse model. The results indicate that tumor tissues grown by the xenografting of PC cells resistant towards docetaxel have a higher disorder strength $L_d$ than the same tissues from drug-sensitive PC cells. Since the disorder strength increases with the increase in the level of tumorigenicity, which implies chemotherapy resistant cells are more aggressive than drug-sensitive cancerous cells. Cells from the prostate or any other cancerous region that survived through drug exposure are more aggressive and develop with the cell line's hierarchy as: PC-3 > DU145 [10]. As an application of the developed finer focusing PWS technique, we have studied the structural alteration in the xenografted tissue morphology that are grown in 3D environment from single cells that mainly have 2D structure. The results show metastasize isolated 2D cancer cells grown to 3D tumors in the body have similar structure and characteristics. The PWS analysis of 3D tissues slices confirms xenografted tissues from drug-resistant tumor have the higher average and standard deviation of disorder strength than the drug-sensitive counter parts. Also, the obtained results follow the similar hierarchy of the cell lines studies.

*Probable cause of structural properties of drug resistance cells and higher structural disorder ($L_d$):* Studies have shown that the progress of cancer disturbs the regular growth as well as the structure of cells/tissues. Further, cells/tissues that survive through the chemotherapy adapt themselves with the situation and develop different morphological structures resulting in higher mass density fluctuation due to the rearrangement of macromolecules, larger pore sizes, changes in cytoskeleton nanoarchitecture, etc. This result in a higher structural alteration in drug-resistant PC cells than drug-sensitive ones. The different mechanisms such as: DNA damage repair, drug inactivation, alteration of drug targets, cancer cells/tissues heterogeneity, cells/tissue death inhibition, epithelial-mesenchymal transition and metastasis, etc. [17–19] are making the cells/tissues drug resistant [10]. Xenografted tumor tissues that are obtained from drug-resistant PC cell lines, which survive due to one of the mechanisms mentioned above, have the same types of structural properties when they are grown into a 3D structure. This supports that the same hierarchy of structural disorder survives in the grown 3D structure.

*Applications of the developed technique for cancer treatment:* The PWS study of xenografted tissues obtained from drug-sensitive and drug-resistant PC cells line could establish a new insight into advancing the understanding of the physical state and drug effectiveness on cancerous cells/tissues at the nanoscale level, by knowing their structural properties. The xenografted tissue structure replicating the structural properties of cancer cells explained statistically in term of the disorder strength $L_d$ parameter could be a reliable, easy, and quantitative approach to diagnose chemo resistance. This result seeks the potential application to monitor the effect of chemotherapy drugs on cancerous tissues and to study the different level of tumorigenicity which can be obtained both, *in-vitro* and *in-vivo* method. In summary, this method will help in understanding the drug-resistant and drug-sensitive cells that are grown within the body, by examining the cells only.

# 5 Acknowledgement

National Institutes of Health (NIH) grants (No. R01EB016983) for Dr. Pradhan. Dr. Yallapu was supported by NIH K22 CA1748841, NIH R15 CA213232.

## 6  References

1. P. Pradhan, D. Damania, H. M. Joshi, V. Turzhitsky, H. Subramanian, H. K. Roy, A. Taflove, V. P. Dravid, and V. Backman, "Quantification of nanoscale density fluctuations using electron microscopy: Light-localization properties of biological cells," Appl. Phys. Lett. **97**(24), 243704 (2010).
2. P. Pradhan, D. Damania, H. M. Joshi, V. Turzhitsky, H. Subramanian, H. K. Roy, A. Taflove, V. Dravid, and V. Backman, "Quantification of Nanoscale Density Fluctuations by Electron Microscopy: probing cellular alterations in early carcinogenesis," Phys Biol **8**(2), 026012 (2011).
3. H. Subramanian, P. Pradhan, Y. Liu, I. R. Capoglu, X. Li, J. D. Rogers, A. Heifetz, D. Kunte, H. K. Roy, A. Taflove, and V. Backman, "Optical methodology for detecting histologically unapparent nanoscale consequences of genetic alterations in biological cells," Proc. Natl. Acad. Sci. U.S.A. **105**(51), 20118–20123 (2008).
4. P. Wang, R. K. Bista, W. E. Khalbuss, W. Qiu, S. Uttam, K. Staton, L. Zhang, T. A. Brentnall, R. E. Brand, and Y. Liu, "Nanoscale nuclear architecture for cancer diagnosis beyond pathology via spatial-domain low-coherence quantitative phase microscopy," J Biomed Opt **15**(6), 066028 (2010).
5. H. Subramanian, P. Pradhan, Y. Liu, I. R. Capoglu, J. D. Rogers, H. K. Roy, R. E. Brand, and V. Backman, "Partial-wave microscopic spectroscopy detects subwavelength refractive index fluctuations: an application to cancer diagnosis," Opt Lett **34**(4), 518–520 (2009).
6. S. Bhandari, P. Shukla, H. Almabadi, P. Sahay, R. Rao, and P. Pradhan, "Optical study of stress hormone-induced nanoscale structural alteration in brain using partial wave spectroscopic (PWS) microscopy," Journal of Biophotonics **0**(ja), e201800002 (n.d.).
7. H. Subramanian, H. K. Roy, P. Pradhan, M. J. Goldberg, J. Muldoon, R. E. Brand, C. Sturgis, T. Hensing, D. Ray, A. Bogojevic, J. Mohammed, J.-S. Chang, and V. Backman, "Nanoscale Cellular Changes in Field Carcinogenesis Detected by Partial Wave Spectroscopy," Cancer Res **69**(13), 5357–5363 (2009).
8. D. Damania, H. Subramanian, A. K. Tiwari, Y. Stypula, D. Kunte, P. Pradhan, H. K. Roy, and V. Backman, "Role of cytoskeleton in controlling the disorder strength of cellular nanoscale architecture," Biophys. J. **99**(3), 989–996 (2010).
9. H. K. Roy, V. Turzhitsky, Y. Kim, M. J. Goldberg, P. Watson, J. D. Rogers, A. J. Gomes, A. Kromine, R. E. Brand, M. Jameel, A. Bogovejic, P. Pradhan, and V. Backman, "Association between rectal optical signatures and colonic neoplasia: potential applications for screening," Cancer Res. **69**(10), 4476–4483 (2009).
12